\documentclass[epj,referee]{svjour}
\usepackage{graphicx}
\usepackage{epsfig}
\usepackage{bm}

\begin{document}

\title{
A Statistical Mechanics Model of Isotropic
Turbulence Well-Defined within the Context of
the $\epsilon$ Expansion
}

\author{Jeong-Man Park$^{1,2}$ and Michael W. Deem$^1$}

\institute{
Department of Physics \& Astronomy, Rice University,
Houston, TX  77005--1892,
\email{mwdeem@rice.edu}
\and
Department of Physics,
The Catholic University of Korea, Puchon 420--743, Korea,
\email{jeong@king.rice.edu}
}
\authorrunning{Park and Deem}
\titlerunning{A Statistical Mechanics Model of Turbulence}

\abstract{
A statistical mechanics model of isotropic turbulence that
renormalizes the effects of turbulent stresses
into a velocity-gradient-dependent random force term
is introduced.
The model is well-defined within the context of the
renormalization group $\epsilon$ expansion, as the effective
expansion parameter is $O(\epsilon)$.  The Kolmogorov constant
and $N$ parameter of turbulence are of order unity, in accord
with experimental results.  Nontrivial 
intermittency corrections to the single-time
structure functions are calculated as a controlled expansion in $\epsilon$.
}


% 47.27.Gs Isotropic turbulence; homogeneous turbulence
% 47.10.+g General theory (see also 83.10 - in rheology)
% 47.27.Ak Fundamentals
% 47.27.Jv High-Reynolds-number turbulence
% 05.10.Cc Renormalization group methods
% 05.60.Cd Classical transport
%47.55.Mh Flows through porous media
%05.60.+w Transport processes: theory

\maketitle


\section{Introduction}

Turbulence remains an outstanding problem in classical mechanics.
Early on, several self-consistent or closure-based statistical models
were introduced, with perhaps Kraichnan's Direct Interaction Approximation
the most well known \cite{Kraichnan1965}.
Within the statistical mechanics literature, attention has focused
upon the random force model that Dominicis and Martin \cite{Martin} 
generalized from  Forster, Nelson, and Stephen's model of a randomly
stirred equilibrium fluid \cite{Forster}. 
This model has received quite a bit of attention, and its 
transport properties have 
been further examined by Yakhot and Orszag and Avellaneda and Majda
and co-workers
\cite{Orszag,Orszag2,Majda2}.
Various treatments, up to and including
field-theoretic $\epsilon$ expansions, have been applied to the
random force model of turbulence.
  As pointed out by Eyink, however, 
even the $\epsilon$ expansion does not lead to a controlled calculation in
the random force model because the term representing the effects of
high Reynolds numbers must be calculated to arbitrarily high order
in perturbation theory, even for an $O(\epsilon)$ calculation
\cite{Eyink}.
More recently, a variety of novel renormalization group techniques
has been applied to the problem of Navier-Stokes turbulence
\cite{Liao1991,Tomassini1997,Esser1999,Giles2001}.

The experimentally observed scaling behavior
of turbulent energy dissipation, often called the Kolmogorov energy cascade,
suggests that there may be a strong analogy between critical phenomena
and turbulence.  Indeed, the search for such an analogy motivated much
of the field-theoretic work on fluid turbulence
\cite{Goldenfeld}.
  Just as there are
wild fluctuations in particle density near a critical point, so to are there
large fluctuations in the fluid velocity at high Reynolds number turbulence.
This similarity suggests that the effects of turbulence may be modeled by
a random, velocity-dependent force, just as the effects of critical
fluctuations can be modeled by a random, density-dependent force in the
standard $\phi^4$ model.  
While there are likely random forces that are independent of the fluid 
velocity in the context of turbulence, there are
also very likely random forces that are velocity dependent.
Thus, the turbulent force should really depend
on the gradient of the velocity, as it is not simply large velocities that
lead to turbulence, but rather regions of high gradient, such as 
walls or boundaries, that lead to turbulent behavior.
Such boundary roughness is one mechanism for breaking the symmetry
from laminar to turbulence fluid flow.

Indeed, random boundary roughness along the walls generates
velocity-gradient-dependent forces in the bulk that
are quenched in time.
Moreover, in practical, engineering-type calculations, the
turbulent forces are often related to 
velocity gradients by a constitutive relation that contains a
``turbulent viscosity'' parameter, the simplest of these
relations being linear \cite{Bird}.
Following this line of reasoning,
 we introduce a new statistical mechanics model  for turbulence
that renormalizes the effect of turbulent stresses into a 
velocity-gradient-dependent term.
This model will turn out to be well-defined within the $\epsilon$ expansion.
That is, the renormalization group theory takes into account all
physical effects of this model, consistently to order $\epsilon$.
Due to its similarity with turbulent viscosity type models,
this approach may lead to a closer connection with practical turbulence
calculations.

We introduce our velocity-gradient-dependent random force model in
Sec.\ 2.  The problem is cast in the framework of
time-dependent field theory in Sec.\ 3. Renormalization
group flow equations are also calculated in this section, and
three appendices provide details of the
renormalization group calculations.
The behavior of
these flow equations in two and three dimensions, and the predictions
for the Kolmogorov energy cascade, are described in Sec.\ 4.
Nontrivial intermittency corrections to the single-time structure functions
are calculated by an operator product expansion in Sec.\ 5.
We conclude in Sec.\ 6.  

\section{Velocity-Gradient-Dependant Random Force Model}

Our goal is to write a form of the Navier-Stokes equation that
contains a random piece, the random piece representing the
statistical effects of the turbulence.  The Navier-Stokes equation
with a random force is
\begin{equation}
\partial_t v_i +
  \sum_k  \textstyle{\prod_{ik}} \sum_j v_j \partial_j v_k
= \nu \nabla^2 v_i + f_i \ ,
\label{1}
\end{equation}
where $\nu = \mu/\rho$ is the kinematic viscosity, and
$f_i = \sum_k \Pi_{ik} (F_k - \partial_k P)/\rho$ is the
 total body force on the fluid.
The presence of the projection operator
 $\hat \Pi_{ik}({\bf k}) = \delta_{ik} - k_i k_k/k^2$
in these formulas ensures that the incompressibility condition
$\nabla \cdot {\bf v}=0$
is maintained \cite{Forster}.  The Fourier transform is defined by
$\hat f({\bf k}) = \int d {\bf x} f({\bf x}) \exp(i {\bf k} \cdot {\bf x})$.

We choose the random force to depend on the gradient of the velocity:
\begin{equation}
f_i({\bf x},t) = - \gamma_{ijk}({\bf x}) \partial_j v_k({\bf x},t) \ .
\label{2}
\end{equation}
We assume that $\gamma_{ijk}$ is symmetric under
exchange of $j$ and $k$, so that the force looks like a turbulent stress.
We average over the statistics of this force using
a field-theoretic representation of
the Navier-Stokes equation. 
We choose the correlation function to be
\begin{equation}
\langle \hat \gamma_{ijk}({\bf k}_1) \hat \gamma_{lmn}({\bf k}_2) \rangle = 
(2 \pi)^d \delta({\bf k}_1+{\bf k}_2) 
\vert {\bf k}_1  \vert^{-y}
D_{ijk}^{lmn} \ ,
\label{3}
\end{equation}
where
\begin{eqnarray}
&& D_{ijk}^{lmn}= 
D_\alpha(
  \delta_{il} \delta_{jm} \delta_{kn}+
  \delta_{in} \delta_{jm} \delta_{kl}
\nonumber \\ && +
  \delta_{il} \delta_{jn} \delta_{km} +
  \delta_{im} \delta_{jl} \delta_{kn} +
  \delta_{im} \delta_{jn} \delta_{kl} +
  \delta_{in} \delta_{jl} \delta_{km}
  )
\nonumber \\ && +
D_\beta (
  \delta_{ij} \delta_{lm} \delta_{kn} +
  \delta_{ik} \delta_{jm} \delta_{ln} +
  \delta_{ik} \delta_{jn} \delta_{lm}+
  \delta_{ij} \delta_{km} \delta_{ln}
    ) \ .
\nonumber \\
\label{4}
\end{eqnarray}
Initially, we treat this as a mathematical problem, taking $y$ to be
arbitrary.  Later, we determine $y$ by requiring that the transport
properties of turbulence are reproduced.
This scaling form of the correlation function, Eq.\ (\ref{3}), applies
only in the inertial, Kolmogorov regime, for wavevectors below an upper
cutoff related to the inverse of the dissipation length scale and above a lower
cutoff related to the inverse of the so-called integral length scale.
It is this Kolmogorov scaling regime that is of interest in the
present work.
Given the form of Eq.\ \ref{2}, the 
parameters $\sqrt D_\alpha$ and $\sqrt D_\beta$
 can be viewed as modeling gradients of the turbulent viscosity.

\section{Renormalization Group Calculations}

We write the Navier-Stokes equation in field-theoretic form so that
the renormalization group can be applied systematically within the
$\epsilon$ expansion \cite{Justin}.  Within the field-theoretic formalism, any
observable can be calculated.  The average velocity, for example,
is given by $v_i({\bf x},t) = \langle b_i({\bf x},t) \rangle$, where the
average over the $b$ field is taken with respect to the 
weight $\exp(-S)$.
Using Eqs.\ \ref{1}--\ref{4},
we arrive at the following action:
\begin{eqnarray}
S &=& \int_{\bf k} \int d t~ \hat { \bar b}_i(-{\bf k},t)
                [ \partial_t + \nu k^2  + \delta(t) ]
                         \hat b_i ({\bf k},t)
\nonumber \\
&&+ i \lambda \int_{{\bf k}_1 {\bf k}_2 {\bf k}_3} \int d t~
                  (2 \pi)^d \delta({\bf k}_1 + {\bf k}_2 + {\bf k}_3)
\nonumber \\ &&~~~~~~~~~~\times
                 k_{1_j} \hat{\bar b}_k^\perp ({\bf k}_1,t)
                          \hat b_k^\perp ({\bf k}_2,t)
                        \hat b_j^\perp({\bf k}_3,t)
\nonumber \\
&&+ \frac{1}{2} \int_{{\bf k}_1 {\bf k}_2 {\bf k}_3 {\bf k}_4} 
                    \int d t_1 d t_2 ~
             (2 \pi)^d \delta({\bf k}_1 + {\bf k}_2 + {\bf k}_3 + {\bf k}_4 )
\nonumber \\ &&~~~~~~~~~~\times
{{\bf k}_2}_j
{{\bf k}_4}_m
D_{ijk}^{lmn} 
\vert {\bf k}_1 + {\bf k}_2\vert^{-y}
\nonumber \\ &&~~~~~~~~~~\times
        \hat {\bar b}_i^\perp ({\bf k}_1, t_1) \hat b_k^\perp ({\bf k}_2, t_1)
    \hat {\bar b}_l^\perp ({\bf k}_3, t_2) \hat b_n^\perp ({\bf k}_4, t_2)
\ .
\nonumber \\
\label{5}
\end{eqnarray}
The notation $\int_{\bf k}$ stands for $\int d {\bf k} / (2 \pi)^d$, the
integrals over time are from $t=0$ to some large time $t=t_f$, and
the summation convention is implied.
This action is written in terms of the divergence-free part of the
field,
 $\hat b_i^\perp({\bf k}) = \sum_k \hat \Pi_{ik}({\bf k}) \hat b_k({\bf k})$.
We have used the Feynman gauge,
adding in a curl-free component in the quadratic terms to 
make later calculations easier.  Initially  $\lambda=1$.
We have used the replica trick \cite{Kravtsov1} to incorporate the
statistical disorder,
but have suppressed these details since they do not enter in a
one-loop calculation.

\begin{figure}[tbp]
\centering
\leavevmode
\includegraphics[height=2in, angle=0]{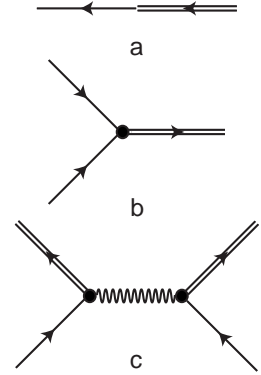}
\caption{(a) Diagram representing the propagator. The arrow points
in the direction of increasing time, and double lines represent the
bar fields. (b) Convection vertex $\lambda$.
(c) Disorder vertex $D_\alpha$ and $D_\beta$.}
\label{fig1}
\end{figure}

We now apply renormalization group theory to this action.
It is important to note that the fields must all have zero average
value before the renormalization group is applied \cite{Nelson1998,Deem2001},
otherwise it would not be correct to truncate perturbation theory 
at any finite order.  
If there were an average velocity, the action in (\ref{5}) would be
different, containing a term of the form
$-i {\bf k} \cdot \langle {\bf v} \rangle$ in the propagator.
  The vertices in the theory are shown in Fig.\ \ref{fig1}.

From power counting,
 the upper critical dimension for this theory is
$d_{\rm c} = 2+y$.
Note that this upper critical dimension is exactly defined
once the model is specified \cite{Justin}.
The deviation of the physical dimension
from the upper critical dimension is parameterized by
$\epsilon = d_{\rm c}-d$.
We use the momentum shell procedure, where fields on a shell of
differential width $\ln a=dl$ are
 integrated out, $\Lambda/a < k < \Lambda$. Note that the combination
$dl$ invariably means a differential on $l$; in all other cases, the
factor $d$ denotes the physical dimension.
As usual, we rescale time by the dynamical exponent
$t' =  a^{-z} t$ and distance 
by ${\bf k}_\perp' = a {\bf k}_\perp $.
The $b$ field is scaled as
$\hat{ b}'({\bf k}',t') = a^{z-1-d + \alpha} \hat{ b}( {\bf k},t)$.
To maintain dimensional consistency, so that the $b$ field
scales as a velocity, one must set $\alpha = 0$ \cite{Deem2001}.
To keep the time derivative in $S$ constant \cite{Deem1},
 the $\bar b$ field  is
scaled as
$\hat{\bar b}'({\bf k}',t') = a^{1-z - \alpha} \hat{\bar b}( {\bf k},t)$.
In the
loop calculation, we make use of the relation for the reference system
averages
\begin{eqnarray}
\langle \hat{ \bar b}_i^\perp ({\bf k}_1,t_1) 
               \hat b_j^\perp ({\bf k}_2, t_2) \rangle_0
&=& (2 \pi)^d \delta({\bf k}_1 + {\bf k}_2) 
\nonumber \\ && \times
       \hat \Pi_{ij}({\bf k}_2) \hat G_0({\bf k}_2, t_2-t_1)
\nonumber \\
\label{5aaa}
\end{eqnarray}
where
$\hat G_0({\bf k},t) = \exp[-\nu k^2 t] \Theta(t)$,
and $\Theta(t) = 1$ if $t > 0$ and 0 otherwise.
Note that elimination of modes at one end of the spectrum
by perturbation theory is the standard procedure in renormalization group
theory \cite{Justin}.  Use of Eq.\ (\ref{5aaa}) does not imply
that the system is somehow Gaussian, as the parameters within the
renormalized theory are flowing.  The critical properties of
the Ising model at the non-Gaussian Wilson-Fisher fixed point,
for example, are analyzed in exactly this way \cite{Justin}.
We make use of the rotational averages:
$\langle k_l k_u \rangle_\Omega = \delta_{lu} k^2 / d$,
$\langle k_l k_m k_s k_u \rangle_\Omega = (
\delta_{lm} \delta_{us}
+
\delta_{lu} \delta_{ms}
+
\delta_{ls} \delta_{mu})
k^4 / [d(d+2)]$, and
$\langle k_l k_m k_n k_s k_t k_u\rangle_\Omega = M_{lmn}^{stu} k^6 /
[d (d+2) (d+4)]$,
where 
the function $M_{lmn}^{stu}$ is equal to all possible couplings of pairs of
the arguments:
\begin{eqnarray}
M_{lmn}^{stu} &=& 
 \delta_{lt} \delta_{ms} \delta_{nu} + 
 \delta_{lt} \delta_{mn} \delta_{su} + 
 \delta_{lt} \delta_{mu} \delta_{ns} 
\nonumber \\  &+&
 \delta_{lm} \delta_{ts} \delta_{nu} + 
 \delta_{lm} \delta_{nt} \delta_{su} + 
 \delta_{lm} \delta_{tu} \delta_{ns}  
\nonumber \\ &+& 
 \delta_{ls} \delta_{mt} \delta_{nu} + 
 \delta_{ls} \delta_{mn} \delta_{tu} + 
 \delta_{ls} \delta_{mu} \delta_{nt}  
\nonumber \\ &+& 
 \delta_{ln} \delta_{mt} \delta_{su} + 
 \delta_{ln} \delta_{ms} \delta_{tu} + 
 \delta_{ln} \delta_{mu} \delta_{st} 
\nonumber \\ &+& 
 \delta_{lu} \delta_{mt} \delta_{ns} + 
 \delta_{lu} \delta_{mn} \delta_{st} + 
 \delta_{lu} \delta_{ms} \delta_{nt} \ .
\label{6}
\end{eqnarray}

\begin{figure}[t]
\centering
\leavevmode
\includegraphics[height=4in, angle=0]{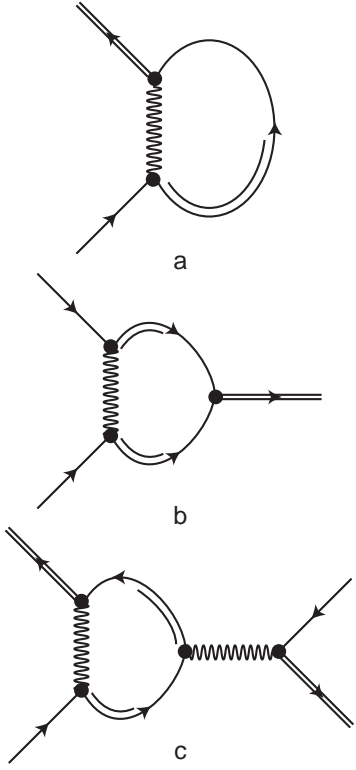}
\caption{One-loop diagrams: (a) self-energy diagrams contributing to
$\nu$, (b) vertex diagrams contributing to $\lambda$, and (c)
vertex diagrams contributing to $D_\alpha$ and $D_\beta$.}
\label{fig2}
\end{figure}

The one loop contributions are shown in Fig.\ \ref{fig2}.
There are  13 diagrams of the form Fig.\ \ref{fig2}a, detailed 
calculation of which is described in Appendix A.
There are 13 similar diagrams of the form Fig.\ \ref{fig2}b,
detailed calculation of which is described in Appendix B.
Finally, there are 100 diagrams of the form Fig.\ \ref{fig2}c,
and a detailed discussion of them is given in Appendix C.
These last diagrams require some care in their calculation, as 
they contribute to the complex tensor structure of
$D_{ijk}^{lmn}$ in Eq.\ (\ref{5}).  These last diagrams
make contributions in the form
\begin{eqnarray}
&&{{\bf k}_2}_j
{{\bf k}_4}_v 
      \hat {\bar b}_i^\perp ({\bf k}_1, t_1) \hat b_k^\perp ({\bf k}_2, t_1)
    \hat {\bar b}_r^\perp ({\bf k}_3, t_2) \hat b_w^\perp ({\bf k}_4, t_2)
\nonumber \\ &&~~~~~~~~~~ \times
M_{lmn}^{stu} 
D_{ijk}^{lmn}
D_{rst}^{uvw} \ .
\end{eqnarray}
There are 15 terms of the form Fig.\ \ref{fig2}c
when the sums over $l,m,n$ and $s,t,u$ are taken.  Each of these
terms corresponds to one of the 15 terms in Eq.\ (\ref{6}), and each gives a 
contribution that is exactly of the form in Eq.\ (\ref{5}).
The theory is, therefore, self-consistent in that no terms are
generated at this order
that are not of the original form, and the
symmetry of the original theory is maintained.

The flow equations that result from the one-loop calculation are
\begin{eqnarray}
\frac{d \ln \nu}{d l} &=& z-2 + [(d^2+2d-2) D_\alpha + (d-2) D_\beta]
\nonumber \\ &&~~~~~~~~~~\times
                             \frac{y K_d}{\nu^2 d (d+2)} \Lambda^{-\epsilon}
\nonumber \\
\frac{d \ln \lambda}{d l} &=& -\alpha -  [(d^2+2d-2) D_\alpha + (d-2) D_\beta]
\nonumber \\ &&~~~~~~~~~~\times
                             \frac{K_d}{\nu^2 d (d+2)} \Lambda^{-\epsilon}
\nonumber \\
\frac{d \ln D_\alpha}{d l} &=& 2z-4 + \epsilon -
      [2 D_\alpha + 4 D_\beta]
\nonumber \\ &&~~~~~~~~~~\times
       \frac{2 K_d (d^2 + 2 d - 2)}{\nu^2 d (d+2) (d+4)} \Lambda^{-\epsilon}
\nonumber \\
\frac{d \ln D_\beta}{d l} &=& 2z-4 + \epsilon -
      [2 (d^2 + 2 d - 2) D_\alpha^2 / D_\beta
\nonumber \\ &&~~~+
        (d^3 + 6 d^2 - 8) D_\alpha
       + (d^2 + 2 d - 8)D_\beta]
\nonumber \\ &&~~~~~~~~~~\times
       \frac{2 K_d }{\nu^2 d (d+2) (d+4)} \Lambda^{-\epsilon} \ .
\label{8a}
\end{eqnarray}
The constant $K_d = S_d / (2 \pi)^d$, and $S_d = 2 \pi^{d/2} / \Gamma(d/2)$.
In solving these equations, we set $\alpha=0$ for dimensional
consistency.  In the standard model of turbulence \cite{Martin}, terms higher
order in $\lambda(l)$ must be kept in the flow equations.  In the
present model, $\lambda(l)$ flows to zero rapidly, and higher order
do not contribute at this level.

\section{Results and Discussion}

In two dimensions, both $D_\alpha$ and $D_\beta$ are relevant.
We find a fixed point of
 $D_\alpha^* = D_\beta^*  = 2 \epsilon \nu^2 \Lambda^\epsilon/ [
3 K_2 (1+y)]$.  The dynamical exponent is given by
$z = 2 - y \epsilon  / [2 (1+y)]$.  The Reynolds number term scales as
$\lambda(l) = \lambda_0 e^{-\epsilon l /[2(1+y)]}$.

In greater than two dimensions, $D_\beta$ reaches a fixed point, but
$D_\alpha$ flows to zero.  We find  
\begin{eqnarray}
D_\beta^* &=& 
    \frac{d (d+2) \nu^2 \Lambda^\epsilon }{2 (1+y) (d-2) K_d} \epsilon
\nonumber \\
z &=& 2 - \frac{y \epsilon}{2 (1+y)}
\nonumber \\
\lambda(l) &=& \lambda_0 e^{-\epsilon l /[2(1+y)]} \ .
\label{8aa}
\end{eqnarray}
Interestingly, the dynamical exponent is the same in 
two and greater dimensions, as is
the decay of $\lambda(l)$.

A particularly beautiful feature of this theory is that $\lambda(l)$
decays exponentially to zero.  This property is what makes the
theory well defined within the epsilon expansion.  If $\lambda(l)$
had stayed at unity, the vertex in Fig.\ \ref{1}b could be inserted
arbitrarily many times in the loop expansion, and terminating the
expansion at any finite order would not be justified by any small
parameter.  The present calculation, on the other hand, is a controlled
expansion in $\epsilon$ and $\lambda(l)$. 
Note that the quenched random forces, which mimic
the effects of, say,  wall roughness,
break statistical Galilean  invariance, and this allows
$\lambda(l)$ to flow, in contrast to the
conventional model with random forces delta-correlated
in time.
Indeed, our explicit calculation shows that $\lambda(l)$ 
decays exponentially to zero.

We now turn to a calculation of the energy spectrum,
defined by  \cite{Orszag1a}
\begin{equation}
E({\bf k}) = \frac{(d-1)}{2} K_d k^{d-1}
 \hat C_{11} ({\bf k}) \ ,
\end{equation}
where
the velocity-velocity correlation function is given in the
field-theoretic language as
\begin{equation}
(2 \pi)^d \delta({\bf k}_1 + {\bf k}_2)  \hat \Pi_{ij}({\bf k}_1)
\hat C_{ij}({\bf k}_1)  =
\langle \hat b_i^\perp({\bf k}_1,t)  \hat b_j^\perp({\bf k}_2,t) \rangle
\ .
\end{equation}
Under the scaling of time and space that occurs within the
renormalization group calculation, this correlation function scales as
\begin{eqnarray}
\hat C_{ij}({\bf k}) &=& \frac{a^{2 (d+1-z - \alpha)}}{a^d} 
  \frac{\langle {\hat{b}_i^\perp}{'} ({\bf k}_1',t')  
   {\hat{b}_j^\perp} {'}({\bf k}_2',t') \rangle}
        {(2 \pi)^d \delta({\bf k}_1 ' + {\bf k}_2 ')\hat\Pi_{ij}({\bf k}_1')}
\nonumber \\
&=&
a^{d+2-2z - 2\alpha} \hat C'_{ij}({\bf k}') \ .
\label{12}
\end{eqnarray}
Making the assumption that 
$\hat C_{ij} ({\bf k}) \sim ({\rm const}) k^{-\delta}$,
we find from Eq.\ \ref{12} that $\delta = d + 2 - 2 z - 2 \alpha$.  The energy
spectrum, therefore, scales as
\begin{equation}
E({\bf k}) \sim ({\rm const}) k^{2z-3 + 2 \alpha} \ .
\label{12a}
\end{equation}
For any isotropic statistical theory of turbulence, then, the
dimensional consistency condition of $\alpha=0$
enforces a relation between the dynamical exponent and the
exponent of the energy cascade.  In particular, the Richardson
separation law implies $z = 2/3$, and this result is equivalent to 
enforcing the Kolmogorov energy cascade:
 $E({\bf k}) \sim ({\rm const}) k^{-5/3}$.
The relation $z=2/3$ implies $y = 3.44152$ in two dimensions and
$y = 4.28849$ in three dimensions.

To calculate the Kolmogorov constant $C_{\rm K}$ we introduce a source
of randomness into the model:
\begin{equation}
\delta S = -\frac{D}{2} \sum_i \int dt \int_{\bf k}
\hat c(k) 
\hat{\bar b}_i^\perp (-{\bf k},t)
\hat{\bar b}_i^\perp({\bf k},t) \ .
\label{14}
\end{equation}
In the range for which scaling occurs, we set $\hat c(k) \sim k^{3z-2-d}$.
For later convenience, we also require that $\lim_{r \to 0} c(r) = 1$.
The randomness expressed in this source term drives the model away from the 
trivial solution $v_i({\bf x},t) \equiv 0$.
Note that the randomness parameter scales as
$D(l) \equiv D$.  This randomness parameter does not contribute to
$\nu$ since $\lambda(l)$ flows to zero, and nothing contributes to 
$D(l)$ at one loop.  Since $D(l)$ contributes to physical properties
at higher loops only through $\lambda(l)$, $D_\alpha(l)$, and $D_\beta(l)$,
all of which are small, the effects of $D$ are controlled within
the $\epsilon$ expansion.
Using the matching Eq.\ \ref{12}, the correlation function is given by
$\hat C_{ij}({\bf k}) = [D \Lambda^{z-2} / (2\nu)] k^{-\delta}$.
For fully-developed isotropic turbulence $z=2/3$, and
using the notation \cite{Orszag}
\begin{equation}
E({\bf k}) \sim C_{\rm K} \varepsilon^{2/3} k^{-5/3} \ ,
\label{15}
\end{equation}
we find $C_{\rm K} \varepsilon^{2/3} =
   (d-1) K_d D \Lambda^{-4/3}/ (4 \nu) $.
Similarly, the wavevector-dependent viscosity considered in the
fluid mechanics literature \cite{Orszag}
is given by 
$\nu({\bf k}) = \nu \Lambda^{4/3} k^{-4/3}$.
Using the notation \cite{Orszag}
\begin{equation}
\nu({\bf k}) \sim N \varepsilon^{1/3} k^{-4/3} \ ,
\end{equation}
we find $N \varepsilon^{1/3} = \nu \Lambda^{4/3}$.
The energy dissipation rate is given by
% \cite{McComb2001}
\begin{equation}
\varepsilon = \int_0^\Lambda dk~ 2 \nu k^2 E(k) \ .
\label{17}
\end{equation}
Using Eqs.\ (\ref{15}) and (\ref{17}), we find
$\varepsilon = (27/8) \nu^3 C_{\rm K}^3 \Lambda^4$.
Finally, to complete the matching we assume that the wavevector
 cutoff is one-half the Kolmogorov dissipation
number,
$2 \Lambda = k_d \equiv (\varepsilon / \nu^3)^{1/4}$
% \cite{McComb2001}. 
\cite{Pao}.
Putting these relations together in three dimensions, we find
\begin{eqnarray}
\varepsilon &=& 0.0380 D
\nonumber \\ 
C_{\rm K} &=& 1.68
\nonumber \\ 
N &=& 0.397 \ .
\end{eqnarray}
Note that field theory cannot calculate non-universal
parameters such as these with precision, as these
results depend on the assumption in Eq.\ (\ref{17})  and the
relation between $\Lambda$ and $k_d$.
A more detailed
matching calculation of these values using the model of turbulence
in Eqs.\ (\ref{3}) and
(\ref{14}) to refine the matching calculation
would be of interest.

\section{Intermittency}
We here address the issue of intermittency in our model.
That is, we seek to determine the scaling of the single-time structure
function 
\begin{equation}
S_{2n}(r) = \langle \left[ 
\vert {\bf b}^\perp ({\bf x} + {\bf r}) - {\bf b}^\perp ({\bf x}) \vert^2
\right]^n \rangle  \sim ({\rm const}) r^{\zeta_{2n}} \ .
\label{18}
\end{equation}
From simple dimensional analysis, we find $\zeta_{2n} = n z$.
From explicit calculation for our model, we find an exponent
that differs from this value.  This difference is referred to
in the fluid mechanics literature as intermittency. 
That $\zeta_{2n} \ne n z$ is an expression of the non-Gaussian nature of
the fixed point identified in our model and of the
divergence of the single-time structure functions in the limit
of an infinite integral length scale.

We use the same arguments about scaling of space and time as in Eq.\
\ref{12} to express the original correlation function in terms of the
renormalized correlation function:
\begin{equation}
S_{2n}(r) \sim e^{n(2-2z)l} S_{2n}(r(l); l) \ .
\label{19}
\end{equation}
Here $r = \exp(l) r(l)$.
This equation is applied until $r(l)$ is of the order of the
dissipation length scale $r(l^*) = 2 \pi / \Lambda \equiv h$.
At this length scale, we then match the correlation function to
a perturbation theory result:
\begin{eqnarray}
S_{2n}(r(l^*); l^*) &\propto& \left[ \frac{h^2}{2 \nu} c(h; l^*) \right]^n
\nonumber \\
&=& \left( \frac{h^2}{2 \nu} \right)^n e^{(3 z - 2)n l^*} c^n(h)
\nonumber \\
&=& \left( \frac{h^2}{2 \nu} \right)^n e^{(3 z - 2)n l^*} \ ,
\label{20}
\end{eqnarray}
where we have used the scaling of the small-$r$ behavior of
the $c(r)$ function in Eq.\ \ref{14} and have used $\lim_{h \to 0} c(h) = 1$.
Combining Eqs.\ \ref{19} and \ref{20} we find
\begin{eqnarray}
S_{2n} (r) \sim ({\rm const}) r^{n z} f_{n} (r/L) \ .
\label{21}
\end{eqnarray}
We have here introduced the fact that traditional renormalization
group arguments can determine asymptotic behavior only up to a scaling
function of the ratio $r/L$, where $L$ is the macroscopic size of the
system. 
This is because $r$ and $L$ are scaled by the same factor in
the renormalization group analysis, and so no dependence on the
ratio $r/L$ is detectable. 
For turbulence, $h$ is the dissipation length scale, and
$L$ is the integral length scale.
In many applications of renormalization
group theory to condensed matter systems, the scaling function
$f_n(r/L) \sim ({\rm const})$ as $L \to \infty$,
and so it does not play a role.
In our case, on the other hand,
the scaling function gives the corrections to intermittency.

To determine the function $f_n$
we use the operator product expansion \cite{Justin,Wilson1969,Kadanoff1969}.
A similar strategy has proven successful
in the study of turbulent transport of passive scalars \cite{Adzhemyan}.
The operator product expansion states that 
\begin{equation}
\langle F({\bf x}_1) F({\bf x}_2) \rangle
\sim \sum_\alpha c_\alpha(\vert {\bf x}_1 - {\bf x}_2 \vert)
\langle
F_\alpha^R({\bf x})
\rangle \ .
\label{22}
\end{equation}
Here the ${\bf x} = ({\bf x}_1 + {\bf x}_2 )/2$,
and $F_\alpha^R$ is the set of all renormalized
operators that are generated by
the renormalization group flow of $F$.
Equation \ref{22} is nothing more than a Taylor series expansion,
where both the bare terms in the Taylor series and those terms
that are generated by the renormalization flow are included.
In our particular case, instead of a pair of operators, 
we have $S_{2n}$, which is a
product of $2n$ factors on the left hand side of this equation.
  The important point about
this expansion is that the functions $c_\alpha(r)$ are
finite and exhibit no dependence on the system size $L$.
Any possible system size dependence of this expansion, therefore, is
contained within $\langle F_\alpha^R({\bf x}) \rangle$.  By comparison
to Eq.\ \ref{21}, we see that the scaling function $f_n$ is thus
determined by the behavior of these renormalized operators.  
We first determine the scaling of the renormalized operators:
\begin{equation}
\langle F_\alpha^R \rangle
=  e^{\Delta_\alpha l^* }
\langle F_\alpha^R(l^*) \rangle
 \ .
\label{23}
\end{equation}
  We 
follow the renormalization flows until $L(l^*) = \exp(-l^*) L = r$,
a criterion that automatically ensures the functional form
specified in Eq.\ \ref{21}.
We thus conclude that
\begin{equation}
\langle F_\alpha^R \rangle
=  \left( \frac{r}{L}\right)^{-\Delta_\alpha}
\langle F_\alpha^R(l^*) \rangle
 \ .
\label{24}
\end{equation}
In fact, we determine the scaling of these operator averages by using the
generating functional $\delta S =
   -A_\alpha \int dt \int d^d {\bf x} F_\alpha^R$.
We will find 
\begin{equation}
A_\alpha(l) \sim e^{(z + d - 2 n z + \gamma_\alpha)l}
A_\alpha(0) \ .
\label{24a}
\end{equation}
In terms of $A_\alpha$ and the partition function $Z$, the
operator average is given by
\begin{equation}
\int dt \int d^d {\bf x} \langle F_\alpha^R \rangle
= \left. \frac{d \ln Z} {d A_\alpha(0)} \right\vert_{A_\alpha(0)=0} \ .
\label{24aa}
\end{equation}
This equation makes clear why Eq.\ \ref{23} has the form that it does
and identifies $\Delta_\alpha = - 2 n z + \gamma_\alpha$.
The value of $\gamma_\alpha$ will be determined by a nontrivial
fixed point of the renormalization group flow equations for
$A_\alpha$.
Combining Eqs.\ \ref{24} and \ref{24a}, we find
\begin{equation}
\langle F_\alpha^R \rangle
 \sim \left(\frac{r}{L}\right)^{2 n z - \gamma_\alpha}
\langle F_\alpha^R(l^*) \rangle \ .
\end{equation}
  The function $\langle F_\alpha^R(l^*) \rangle$
 is determined by matching exactly
as in Eq.\  \ref{20}.  We, thus, find that
\begin{eqnarray}
\langle F_\alpha^R \rangle &\sim &
\left(\frac{r}{L}\right)^{2 n z - \gamma_\alpha}
\left(\frac{r}{L}\right)^{(2 - 3 z)n} 
\nonumber \\
&=&
\left(\frac{r}{L}\right)^{n(2-z) - \gamma_\alpha} \ .
\label{25}
\end{eqnarray}
To make use of this result of the operator product expansion,
it remains only to calculate the value of $\gamma_\alpha$.
Once we have this value, we find the scaling function to be
\begin{equation}
f_{2n}(r/L) \sim 
\left(\frac{r}{L}\right)^{n(2-z) - \gamma_\alpha} 
{\rm as~}L{\rm ~to~} \infty \ .
\label{26}
\end{equation}
In fact, the operator $S_{2n}$ will generate several new
operators $F_\alpha$ in the expansion of Eq.\ \ref{22}.
The appropriate value of $\gamma_\alpha$ to use in Eq.\ \ref{26}
is the largest one.  These generated operators may mix
upon the renormalization, in which case the appropriate value
of $\gamma_\alpha$ is the largest eigenvalue  of the
flow equation matrix.

We first determine the scaling function $f_1$.  We limit consideration to the
case $d > 2$, where $D_\alpha^* = 0$.  The correlation
function $S_{2n}$ is given in a Taylor series as $S_{2n}(r) \sim
[r^2 \partial_x b_i \partial_x b_i]^n$.  This will generate
the symmetrized operator
$F = [\partial_j b_i \partial_j b_i + \partial_i b_j \partial_j b_i]^n$,
which we consider.

For the case $n=1$, we consider the generating functional
\begin{eqnarray}
\delta S_{II} &=&
-\int dt \int_{{\bf k}_1 {\bf k}_2}
\bigg[
 A^{(1)} {\bf k}_1 \cdot {\bf k}_2
\hat{ b}^\perp ({\bf k}_1,t) \cdot \hat{ b}^\perp({\bf k}_2,t)
\nonumber \\
&&+
A^{(0)} {\bf k}_1 \cdot \hat{ b}^\perp ({\bf k}_2,t)
{\bf k}_2 \cdot \hat{ b}^\perp({\bf k}_1,t)
\bigg]
 \ .
\label{27}
\end{eqnarray}
We find
\begin{equation}
\frac{d  A^{(i)}}{dl} = (z+d-2z) A^{(i)} + \frac {2 (d-1) g}{d} A^{(1)} \  ,
\label{28}
\end{equation}
where $g = D_\beta K_d / (\nu^2 \Lambda^\epsilon)$.
We, thus, identify $\gamma_1 = 2 (d-1) g^*/d$.
Using Eq.\ \ref{8aa} we find in three dimensions
$z = 2 - \epsilon/4 + O(\epsilon^2)$ and
$g^* = 15 \epsilon/4 + O(\epsilon^2)$
 and conclude from Eqs.\ \ref{21} and \ref{26}
that
\begin{eqnarray}
S_2(r) &\sim& ({\rm const}) r^z \left( \frac{r}{L} \right)^{2-z - \gamma_1}
\nonumber \\
&=&  ({\rm const}) r^{2-\epsilon/4}
   \left( \frac{r}{L} \right)^{\epsilon/4-5 \epsilon} \ .
\label{29}
\end{eqnarray}

We now determine the scaling function $f_2$.
We start with the symmetrized generating functional
\begin{eqnarray}
\delta S^{(2)} &=& 
-\int dt \int_{{\bf k}_1 {\bf k}_2 {\bf k}_3 {\bf k}_4}
\hat{ b}^\perp ({\bf k}_1,t) \cdot \hat{ b}^\perp({\bf k}_2,t)
\nonumber \\
&&\times
\hat{ b}^\perp ({\bf k}_3,t) \cdot \hat{ b}^\perp({\bf k}_4,t)
\bigg[
 A_1^{(2)} {\bf k}_1 \cdot {\bf k}_2 {\bf k}_3 \cdot {\bf k}_4
\nonumber \\
&& +
 A_2^{(2)} ({\bf k}_1 \cdot {\bf k}_3 {\bf k}_2 \cdot {\bf k}_4
+ {\bf k}_1 \cdot {\bf k}_4 {\bf k}_2 \cdot {\bf k}_3)
\bigg] \ .
\end{eqnarray}
This term generates two additional generating functionals:
\begin{eqnarray}
\delta S^{(1)} &=&
-\int dt \int_{{\bf k}_1 {\bf k}_2 {\bf k}_3 {\bf k}_4}
\hat{ b}^\perp ({\bf k}_1,t) \cdot \hat{ b}^\perp({\bf k}_2,t)
\nonumber \\ &&\times
\bigg[
A_1^{(1)} {\bf k}_1 \cdot {\bf k}_2 
       {\bf k}_4 \cdot \hat{ b}^\perp({\bf k}_3,t)
       {\bf k}_3 \cdot \hat{ b}^\perp({\bf k}_4,t)
\nonumber \\ &&
+
A_2^{(1)} \bigg(
{\bf k}_1 \cdot {\bf k}_3
{\bf k}_4 \cdot \hat{ b}^\perp({\bf k}_3,t)
{\bf k}_2 \cdot \hat{ b}^\perp({\bf k}_4,t)
\nonumber \\ && +
{\bf k}_1 \cdot {\bf k}_4
{\bf k}_2 \cdot \hat{ b}^\perp({\bf k}_3,t)
{\bf k}_3 \cdot \hat{ b}^\perp({\bf k}_4,t)
\nonumber \\ && +
{\bf k}_2 \cdot {\bf k}_3
{\bf k}_4 \cdot \hat{ b}^\perp({\bf k}_3,t)
{\bf k}_1 \cdot \hat{ b}^\perp({\bf k}_4,t)
\nonumber \\ && +
{\bf k}_2 \cdot {\bf k}_4
{\bf k}_1 \cdot \hat{ b}^\perp({\bf k}_3,t)
{\bf k}_3 \cdot \hat{ b}^\perp({\bf k}_4,t)
\bigg)
\nonumber \\ && +
A_3^{(1)} \bigg(
{\bf k}_3 \cdot {\bf k}_4
{\bf k}_1 \cdot \hat{ b}^\perp({\bf k}_3,t)
{\bf k}_2 \cdot \hat{ b}^\perp({\bf k}_4,t)
\nonumber \\ && +
{\bf k}_3 \cdot {\bf k}_4
{\bf k}_2 \cdot \hat{ b}^\perp({\bf k}_3,t)
{\bf k}_1 \cdot \hat{ b}^\perp({\bf k}_4,t)
\bigg)
\bigg]
\end{eqnarray}
and
\begin{eqnarray}
\delta S^{(0)} &=&
-\int dt \int_{{\bf k}_1 {\bf k}_2 {\bf k}_3 {\bf k}_4}
\hat{ b}_i^\perp ({\bf k}_1,t)  \hat{ b}_j^\perp({\bf k}_2,t)
\hat{ b}_k^\perp ({\bf k}_3,t)  \hat{ b}_l^\perp({\bf k}_4,t)
\nonumber \\ &&\times
\bigg[
A_1^{(0)} {{\bf k}_1}_j {{\bf k}_2}_i {{\bf k}_3}_l {{\bf k}_4}_k
\nonumber \\ &&+
A_2^{(0)} \bigg(
{{\bf k}_1}_j {{\bf k}_2}_l {{\bf k}_3}_i {{\bf k}_4}_k
+
{{\bf k}_1}_j {{\bf k}_2}_k {{\bf k}_3}_l {{\bf k}_4}_i
\nonumber \\ &&+
{{\bf k}_1}_l {{\bf k}_2}_i {{\bf k}_3}_j {{\bf k}_4}_k
+
{{\bf k}_1}_k {{\bf k}_2}_i {{\bf k}_3}_l {{\bf k}_4}_j
\bigg)
\nonumber \\ &&+
A_3^{(0)} \bigg(
{{\bf k}_1}_k {{\bf k}_2}_l {{\bf k}_3}_i {{\bf k}_4}_j
+
{{\bf k}_1}_k {{\bf k}_2}_l {{\bf k}_3}_j {{\bf k}_4}_i
\nonumber \\ &&+
{{\bf k}_1}_l {{\bf k}_2}_k {{\bf k}_3}_i {{\bf k}_4}_j
+
{{\bf k}_1}_l {{\bf k}_2}_k {{\bf k}_3}_j {{\bf k}_4}_i
\bigg)
\bigg]
\end{eqnarray}
We will find $A_2^{(1)}(l) = A_3^{(1)}(l)$.
Although we have included
$A_3^{(0)}$ for generality, we will find that this term is not
generated, and $A_3^{(0)}(l) = 0$.

A lengthy calculation shows that 
\begin{equation}
\frac{d {\bf A}}{ d l} = (z+d-4 z)I {\bf A} +
\frac{g}{d (d+2) (d+4)} M {\bf A}
\label{40}
\end{equation}
where the vector ${\bf A} = (A_1^{(2)}, A_2^{(2)}, A_1^{(1)}, A_2^{(1)},
A_3^{(1)}, A_1^{(0)}, A_2^{(0)} )$, and
the matrix $M$ is given in Table \ref{matrixM}.
\begin{table*}[p]
\begin{minipage}[h]{6.00in}
\caption{The matrix $M$ in the calculation of the intermittency
exponent $\zeta_4$ via Eq.\ \ref{40}.}
\label{matrixM}
\begin{eqnarray*}
M = 
\left( \begin{array}{ccccccc}
4 (d^3 + 5 d^2 + 2 d - 6)& 8 (d+3)^2& 8&16&4 (d^2 + 6 d + 10)& 8&16 \\
2 (d+2)^2& 2 (d^3 + 5 d^2 - 12) & -4 (d+2)& -8(d+2) &d^3 + 4 d^2 - 8 d - 24
                      &2(d+2)^2&4 (d+2)^2 \\
4 (d^3 + 5 d^2 + 2 d - 4)& 8 (d^2 + 6 d + 10)& 2 (d^3 + 5 d^2 + 2 d)& 
       8 (d^2 + 6 d + 12)&
                     4(d^2 + 6 d + 12) & 16 & 32 \\
2 d (d+2) & 2 (d^3 + 5 d^2 - 2 d - 16) & 2 d (d+2) &
    2(d^3 + 5 d^2 - 6 d -  24) & d^3 + 4 d^2 - 12 d - 32 & 2 d (d +2)&
    4 d (d +2) \\
2d (d +2) & 2 (d^3 + 5 d^2 - 2 d - 16) & 2 d (d +2) & 
   2 ( d^3 + 5 d^2 - 6 d - 24) & d^3 + 4 d^2 - 12 d - 32 &
   2 d (d +2 ) & 4 d (d +2) \\
8 & 8 & 2 (d^3 + 5 d^2 + 2 d - 4) & 8 (d^2 + 6 d + 10) & 8 & 8 & 16 \\
-2 (d+2) & -2 (d+2) & (d+2)^2 & d^3 + 5 d^2 - 2 d - 16 & 
   -2 (d+2) & -2 (d+2) & -4 (d+2) 
\end{array} \right)
\end{eqnarray*}
\end{minipage}
\end{table*}
We diagonalize the matrix $M$, finding the eigenvalues
$(0,0,0,0, d^3 + 4 d^2 - 4 d - 16,
2(d^3 + 5 d^2 + 2 d - 8),
4 (d^3 + 7 d^2 + 6 d) )$.
The largest eigenvalue is the last one for all $d$.
In three dimensions we identify $\gamma_2 = 432 g^* / 105$.
Using Eqs.\  \ref{21} and \ref{26}
we find that
\begin{equation}
S_4(r) \sim
 ({\rm const}) r^{4-\epsilon/2}
   \left( \frac{r}{L} \right)^{\epsilon/2-108 \epsilon/7} \ .
\end{equation}

We have, therefore, derived the intermittency corrections to dimensional
analysis for the present model.  The corrections are calculated in
a controlled fashion and are proportional to $\epsilon$.  The
coefficients of the correction are not small, and for finite $\epsilon$,
higher order terms in the expansion are required for an accurate
estimation of the effects of intermittency.

One might wonder whether there are any 
corrections to the Kolmogorov energy cascade, Eq.\  \ref{12a},
that arise from the operator product expansion.  More generally,
are there any corrections to $\langle [b_i^\perp ({\bf x}, t) 
b_i^\perp ({\bf x}, t)]^n \rangle$?
There are no such corrections.  This type of operator flows under the
renormalization group only to operators with more derivatives, such as
$ \partial_l b_j^\perp \partial_m b_k^\perp (b_i^\perp b_i^\perp )^{n-1}$.
These operators are less relevant than the original operator, and so they
make no contribution to the scaling at leading order.
In the language of Eq.\ \ref{22}, $F_\alpha^R \equiv 1$, and the
scaling function $f_{n}(r/L) \sim ({\rm const})$ as $L \to \infty$.

\section{Conclusion}
An alternative, simpler model would have been to take the turbulent
forces to be proportional to the velocity, rather than the velocity gradient.
The simplest model, moreover, would take the forces to be
white noise in time and uncorrelated in each of the spatial dimensions.
For this model to be nontrivial, a
mean fluid flow must be introduced \cite{Deem2001}.
Interestingly, when this is done for forces that are random in time as
well as space,
the resulting theory has the same flow equations as the
traditional random force model of turbulence \cite{Martin,Forster}.
Also of interest to note is that
a theory with velocity-gradient-dependent random forces that
are white noise in time would have no renormalization of any parameter, as
the diagrams of Fig.\ \ref{fig2} would all vanish due to the causality
of the bare propagator,  Eq.\ (\ref{5aaa}).

The scaling of the Kolmogorov energy cascade is determined once
the value of the dynamical exponent is fixed, \emph{i.e.}\ $z=2/3$
for isotropic turbulence.  In random force models such as the
present one, the scaling of the energy cascade 
simply serves to fix the correlation function of the random forcing.
The predictive power of models such as these lie in their
ability to provide nontrivial predictions of the intermittency
corrections.  In the present model, we are able to provide these
corrections as a systematic expansion in $\epsilon$.

In summary, we have introduced a new statistical mechanics model for isotropic
turbulence.  This model makes use of a random, velocity-gradient-dependent
force.  This model is both consistent with practical, engineering-type
calculations and well-defined within the renormalization group
$\epsilon$ expansion.  This model makes stronger the analogy between
turbulence and critical phenomena.

Owing to the irrelevance of the convection terms at the fixed
point, our results may alternatively be viewed  as an analysis of transport
in a new class of random media.

\begin{acknowledgement}
This research was supported by the National Science Foundation and
by an Alfred P. Sloan Foundation Fellowship to M.W.D.
\end{acknowledgement}

\section*{Appendix A: One-Loop contributions to $\nu$}
We here show how the diagrams of Fig.\ \ref{fig2}a 
contribute to the propagator, Fig.\ \ref{fig1}a.  
Each of the terms is associated
with one of the $D_{ijk}^{lmn}$ terms in
Eq.\ (\ref{5}).
 In the
calculation of the averages on the shell, there are five
terms associated with $D_\alpha$ and three associated with
$D_\beta$, as the last two terms associated with $D_\alpha$ are
identical, and the last two terms associated with $D_\beta$ are
also identical.
The first term, associated with $D_\alpha$, is
\begin{eqnarray}
I_1 &=& 2 \times \frac{D_\alpha}{2} \int d t_1 d t_2 \int_{
{\bf k}_1 {\bf k}_2 {\bf k}_3 {\bf k}_4} 
\nonumber \\ && \times
(2 \pi)^d \delta({\bf k}_1+ {\bf k}_2 + {\bf k}_3 + {\bf k}_4 )
\vert {\bf k}_1 + {\bf k}_2 \vert^{-y}
\nonumber \\ && \times
 \Theta (t_1 - t_2)
(2 \pi)^d \delta({\bf k}_2 + {\bf k}_3)
e^{-\nu k_2^2 (t_1 - t_2)}
\nonumber \\ && \times
\left\{ {\bf k}_2 \cdot {\bf k}_4
\hat{ \bar b}_i^\perp({\bf k}_1,t_1) \hat b_j^\perp({\bf k}_4,t_2)
\hat \Pi_{ij} ({\bf k}_2)
\right\}
\nonumber \\
&=& \frac{K_d}{\nu^2}
   \left[ D_\alpha \frac{y(d+1) }{ d (d+2)} \right]
     \int_{\Lambda/a}^\Lambda dq  q^{d-y-3}
\nonumber \\ && \times
\int dt \int_{\bf k} \nu k^2 \hat{\bar b}_i^\perp(-{\bf k},t) \hat b_i^\perp({\bf k},t)
\label{eqI1}
\end{eqnarray}
For the remaining contributions, we list the symmetry factor,
terms in braces in the integrand of Eq.\ (\ref{eqI1}) that change,
and the final contribution in brackets that change.
The contributions are shown in Table \ref{intI}, where
the dependence of the fields upon time has been suppressed.
The term $e^{-\nu k_i^2(t_1-t_2)}$, which has the same
momentum argument as the $\hat \Pi({\bf k}_i)$ term, has been
suppressed.  The delta function is also suppressed.
Summing all these contributions to $\nu$, we get the first
flow equation of Eq.\ (\ref{8a}).
\begin{table*}[p]
\begin{minipage}[h]{6.00in}
\caption{Terms in 
the contributions from Fig.\ \ref{fig2}a, via Eq.\ \ref{eqI1}.}
\label{intI}
\begin{tabular}{llcc}
\hline
Integral & Symmetry & Integrand & Result\\
\hline
$I_1$ & 2 &
$
\left\{ { k}_{2_l}  { k}_{4_l}
\hat{ \bar b}_i^\perp({\bf k}_1) \hat b_j^\perp({\bf k}_4)
\hat \Pi_{ij} ({\bf k}_2)
\right\} 
$
&
$\left[ D_\alpha \frac{y(d+1) }{ d (d+2) } \right]$
\\
$I_2$ & 2 &
$\left\{ { k}_{2_l}  { k}_{4_l}
\hat{ \bar b}_i^\perp({\bf k}_1) \hat b_i^\perp({\bf k}_4)
\hat \Pi_{jj} ({\bf k}_2)
\right\} $
&
$\left[ D_\alpha \frac{y(d-1) }{ d } \right]$
\\
$I_3^a$ & 1 &
$\left\{ { k}_{2_i} { k}_{4_j}
\hat{ \bar b}_k^\perp({\bf k}_1) \hat b_i^\perp({\bf k}_4)
\hat \Pi_{jk} ({\bf k}_2)
\right\} $
&
$\left[ -\frac{D_\alpha}{2} \frac{y }{  d(d+2) } \right]$
\\
$I_3^b$ & 1 &
$\left\{ { k}_{2_i} { k}_{4_j}
\hat{ \bar b}_k^\perp({\bf k}_3) \hat b_j^\perp({\bf k}_2)
\hat \Pi_{ik} ({\bf k}_4)
\right\} $
&
$\left[ -\frac{D_\alpha}{2} \frac{y }{  d(d+2) } \right]$
\\
$I_4^a$ & 1 &
$\left\{ { k}_{2_i}  { k}_{4_j}
\hat{ \bar b}_j^\perp ({\bf k}_1) \hat b_k^\perp({\bf k}_4)
\hat \Pi_{ik} ({\bf k}_2)
\right\} $
&
0
\\
$I_4^b$ & 1 &
$\left\{ { k}_{2_i}  { k}_{4_j}
\hat{ \bar b}_i^\perp ({\bf k}_3) \hat b_k^\perp({\bf k}_2)
\hat \Pi_{jk} ({\bf k}_4)
\right\} $
&
0
\\
$I_5^a$ & $1 \times 2$ &
$\left\{ { k}_{2_i}  { k}_{4_j}
\hat{ \bar b}_j^\perp ({\bf k}_1) \hat b_i^\perp({\bf k}_4)
\hat \Pi_{kk} ({\bf k}_2)
\right\} $
&
0
\\
$I_5^b$ & $1 \times 2$ &
$\left\{ { k}_{2_i}  { k}_{4_j}
\hat{ \bar b}_k^\perp ({\bf k}_3) \hat b_k^\perp({\bf k}_2)
\hat \Pi_{ij} ({\bf k}_4)
\right\} $
&
0
\\
$I_6^a$ & $1 $ &
$\left\{ { k}_{2_i}  { k}_{4_j}
\hat{ \bar b}_i^\perp ({\bf k}_1) \hat b_k^\perp({\bf k}_4)
\hat \Pi_{jk} ({\bf k}_2)
\right\} $
&
$\left[ -\frac{D_\beta}{2} \frac{y }{  d(d+2) } \right]$
\\
$I_6^b$ & 1  &
$\left\{ { k}_{2_i}  { k}_{4_j}
\hat{ \bar b}_j^\perp ({\bf k}_3) \hat b_k^\perp({\bf k}_2)
\hat \Pi_{ik} ({\bf k}_4)
\right\} $
&
$\left[ -\frac{D_\beta}{2} \frac{y }{  d(d+2) } \right]$
\\
$I_7$ & 2  &
$\left\{ { k}_{2_l}  { k}_{4_l}
\hat{ \bar b}_i^\perp ({\bf k}_1) \hat b_j^\perp({\bf k}_4)
\hat \Pi_{ij} ({\bf k}_2)
\right\} $
&
$\left[ D_\beta \frac{y (d+1) }{ d(d+2) } \right]$
\\
$I_8^a$ & $1 \times 2$  &
$\left\{ { k}_{2_i}  { k}_{4_j}
\hat{ \bar b}_i^\perp ({\bf k}_1) \hat b_k^\perp({\bf k}_4)
\hat \Pi_{jk} ({\bf k}_2)
\right\} $
&
$\left[ -D_\beta \frac{y }{ d(d+2) } \right]$
\\
$I_8^b$ & $1 \times 2$  &
$\left\{ { k}_{2_i}  { k}_{4_j}
\hat{ \bar b}_k^\perp ({\bf k}_3) \hat b_j^\perp({\bf k}_2)
\hat \Pi_{ik} ({\bf k}_4)
\right\} $
&
$\left[ -D_\beta \frac{y }{ d(d+2) } \right]$
\end{tabular}
\end{minipage}
\end{table*}

\section*{Appendix B: One-Loop contributions to $\lambda$}
We here show how the diagrams of Fig.\ \ref{fig2}b 
contribute to the convection term, Fig.\ \ref{fig1}b.  
Each of the terms is associated
with one of the $D_{ijk}^{lmn}$ terms in
Eq.\ (\ref{5}).
It is convenient to define the convection operator
$\hat M_{ijk}({\bf k}) = k_j \hat \Pi_{ik}({\bf k}) +
                         k_k \hat \Pi_{ij}({\bf k})$.
The convection term of Eq.\ (\ref{5}) then becomes
\begin{eqnarray}
&+& \frac{i \lambda}{2} \int_{{\bf k}_1 {\bf k}_2 {\bf k}_3} \int d t~
                  (2 \pi)^d \delta({\bf k}_1 + {\bf k}_2 + {\bf k}_3)
\nonumber \\ && \times
                 \hat M_{ijk}( k_1) \hat{\bar b}_i^\perp ({\bf k}_1,t)
                          \hat b_j^\perp ({\bf k}_2,t)
                        \hat b_k^\perp({\bf k}_3,t) \ .
\end{eqnarray}
 In the
calculation of the averages on the shell, there are again five
terms associated with $D_\alpha$ and three associated with
$D_\beta$.
The first such term is
\begin{eqnarray}
J_1 &=& 2 \times \left(-\frac{1}{2!}\right) 2
 \left( \frac{i \lambda}{2} \right)
\frac{D_\alpha}{2}
\int d t_1 d t_2 
\nonumber \\ && \times
\int_{{\bf k}_1 {\bf k}_2 {\bf k}_3 {\bf k}_4} 
(2 \pi)^d \delta({\bf k}_1+ {\bf k}_2 + {\bf k}_3 + {\bf k}_4 )
\nonumber \\ && \times
\int d t_3
\int_{{\bf k}_5 {\bf k}_6 {\bf k}_7}
(2 \pi)^d \delta({\bf k}_5+ {\bf k}_6 + {\bf k}_7)
\nonumber \\ && \times
\vert {\bf k}_1 + {\bf k}_2 \vert^{-y}
 \Theta (t_1 - t_3)
 \Theta (t_3 - t_2)
\nonumber \\ && \times
(2 \pi)^d \delta({\bf k}_2 + {\bf k}_5)
(2 \pi)^d \delta({\bf k}_6 + {\bf k}_3)
\nonumber \\ && \times
e^{-\nu k_2^2 (t_1 - t_3)}
e^{-\nu k_6^2 (t_3 - t_2)}
\hat M_{lmn}({\bf k}_5)
\nonumber \\ && \times
\bigg\{ 
k_{2_k} k_{4_k}
\hat{ \bar b}_i^\perp({\bf k}_1,t_1) \hat b_j^\perp({\bf k}_4,t_2) \hat b_n^\perp({\bf k}_7,t_3)
\nonumber \\ && \times
\hat \Pi_{jl}({\bf k}_2)
\hat \Pi_{im}({\bf k}_6)
\bigg\}
\nonumber \\
&=& \frac{i \lambda}{2}
\frac{K_d}{\nu^2}
 \left[ - D_\alpha  \frac{(d+1) }{ d (d+2)} \right]
     \int_{\Lambda/a}^\Lambda dq  q^{d-y-3}
\nonumber \\ && \times
\int dt \int_{{\bf k}_1 {\bf k}_2 {\bf k}_3}
(2 \pi)^d \delta({\bf k}_1 + {\bf k}_2 + {\bf k}_3)
\nonumber \\ && \times
\hat M_{ijk}({\bf k}_1)
\hat{\bar b}_i^\perp({\bf k}_1,t) \hat b_j^\perp({\bf k}_2,t) \hat b_k^\perp({\bf k}_3,t)
\label{eqJ1}
\end{eqnarray}
For the remaining contributions, we list the symmetry factor,
terms in braces in the integrand of Eq.\ (\ref{eqJ1}) that change,
and the final contribution in brackets that change.
The contributions are shown in Table \ref{intJ}, where
again the dependence on time has been suppressed.
The terms $e^{-\nu k_i^2(t_1-t_3)}$ and
$e^{-\nu k_j^2(t_3-t_2)}$, which have the same
momentum arguments as the two $\hat \Pi({\bf k}_i)$ terms, have been
suppressed.
The delta functions have also been suppressed.
Summing all these contributions to $\lambda$, we get the second
flow equation of Eq.\ (\ref{8a}).
\begin{table*}[p]
\begin{minipage}[h]{6.00in}
\caption{Terms in
the contributions from Fig.\ \ref{fig2}b, via Eq.\ \ref{eqJ1}.}
\label{intJ}
\begin{tabular}{llcc}
\hline
Integral & Symmetry & Integrand & Result\\
\hline
$J_1$ & 2 &
$
\left\{ { k}_{2_k}  { k}_{4_k}
\hat{ \bar b}_i^\perp({\bf k}_1) \hat b_j^\perp({\bf k}_4) \hat b_n^\perp({\bf k}_7)
\hat \Pi_{jl}({\bf k}_2) \hat \Pi_{im}({\bf k}_6)
\right\} 
$
&
$\left[ - D_\alpha  \frac{(d+1) }{ d (d+2)} \right]$
\\
$J_2$ & 2 &
$
\left\{ { k}_{2_k}  { k}_{4_k}
\hat{ \bar b}_i^\perp({\bf k}_1) \hat b_i^\perp({\bf k}_4) \hat b_n^\perp({\bf k}_7)
\hat \Pi_{jl}({\bf k}_2) \hat \Pi_{mj}({\bf k}_6)
\right\} 
$
&
$\left[ - D_\alpha  \frac{(d-1) }{ d } \right]$
\\
$J_3^a$ & 1 &
$
\left\{ { k}_{2_i}  { k}_{4_j}
\hat{ \bar b}_k^\perp({\bf k}_1) \hat b_i^\perp({\bf k}_4) \hat b_n^\perp({\bf k}_7)
\hat \Pi_{jl}({\bf k}_2) \hat \Pi_{km}({\bf k}_6)
\right\} 
$
&
$\left[ + \frac{D_\alpha}{2}  \frac{1 }{ d (d+2)} \right]$
\\
$J_3^b$ & 1 &
$
\left\{ { k}_{2_i}  { k}_{4_j}
\hat{ \bar b}_k^\perp({\bf k}_3) \hat b_j^\perp({\bf k}_2) \hat b_n^\perp({\bf k}_7)
\hat \Pi_{il}({\bf k}_4) \hat \Pi_{km}({\bf k}_6)
\right\} 
$
&
$\left[ + \frac{D_\alpha}{2}  \frac{ 1}{ d (d+2)} \right]$
\\
$J_4^a$ & 1 &
$
\left\{ { k}_{2_i}  { k}_{4_j}
\hat{ \bar b}_j^\perp({\bf k}_1) \hat b_k^\perp({\bf k}_4) \hat b_n^\perp({\bf k}_7)
\hat \Pi_{lk}({\bf k}_2) \hat \Pi_{im}({\bf k}_6)
\right\} 
$
&
0
\\
$J_4^b$ & 1 &
$
\left\{ { k}_{2_i}  { k}_{4_j}
\hat{ \bar b}_i^\perp({\bf k}_3) \hat b_k^\perp({\bf k}_2) \hat b_n^\perp({\bf k}_7)
\hat \Pi_{kl}({\bf k}_4) \hat \Pi_{mj}({\bf k}_6)
\right\} 
$
&
0
\\
$J_5^a$ & $1 \times 2$ &
$
\left\{ { k}_{2_i}  { k}_{4_j}
\hat{ \bar b}_j^\perp({\bf k}_1) \hat b_i^\perp({\bf k}_4) \hat b_n^\perp({\bf k}_7)
\hat \Pi_{kl}({\bf k}_2) \hat \Pi_{mk}({\bf k}_6)
\right\} 
$
&
0
\\
$J_5^b$ & $1 \times 2$ &
$
\left\{ { k}_{2_i}  { k}_{4_j}
\hat{ \bar b}_k^\perp({\bf k}_3) \hat b_k^\perp({\bf k}_2) \hat b_n^\perp({\bf k}_7)
\hat \Pi_{il}({\bf k}_4) \hat \Pi_{mj}({\bf k}_6)
\right\} 
$
&
0
\\
$J_6^a$ & 1 &
$
\left\{ { k}_{2_i}  { k}_{4_j}
\hat{ \bar b}_i^\perp({\bf k}_1) \hat b_k^\perp({\bf k}_4) \hat b_n^\perp({\bf k}_7)
\hat \Pi_{kl}({\bf k}_2) \hat \Pi_{mj}({\bf k}_6)
\right\} 
$
&
$\left[ + \frac{D_\beta}{2}  \frac{ 1}{ d (d+2)} \right]$
\\
$J_6^b$ & 1 &
$
\left\{ { k}_{2_i}  { k}_{4_j}
\hat{ \bar b}_j^\perp({\bf k}_3) \hat b_k^\perp({\bf k}_2) \hat b_n^\perp({\bf k}_7)
\hat \Pi_{kl}({\bf k}_4) \hat \Pi_{im}({\bf k}_6)
\right\} 
$
&
$\left[ + \frac{D_\beta}{2}  \frac{ 1}{ d (d+2)} \right]$
\\
$J_7$ & 2 &
$
\left\{ { k}_{2_k}  { k}_{4_k}
\hat{ \bar b}_i^\perp({\bf k}_1) \hat b_j^\perp({\bf k}_4) \hat b_n^\perp({\bf k}_7)
\hat \Pi_{il}({\bf k}_2) \hat \Pi_{jm}({\bf k}_6)
\right\} 
$
&
$\left[ - D_\beta  \frac{ (d+1)}{ d (d+2)} \right]$
\\
$J_8^a$ & $1 \times 2$ &
$
\left\{ { k}_{2_i}  { k}_{4_j}
\hat{ \bar b}_i^\perp({\bf k}_1) \hat b_k^\perp({\bf k}_4) \hat b_n^\perp({\bf k}_7)
\hat \Pi_{jl}({\bf k}_2) \hat \Pi_{km}({\bf k}_6)
\right\} 
$
&
$\left[ + D_\beta  \frac{ 1}{ d (d+2)} \right]$
\\
$J_8^b$ & $1 \times 2$ &
$
\left\{ { k}_{2_i}  { k}_{4_j}
\hat{ \bar b}_k^\perp({\bf k}_3) \hat b_j^\perp({\bf k}_2) \hat b_n^\perp({\bf k}_7)
\hat \Pi_{kl}({\bf k}_4) \hat \Pi_{mi}({\bf k}_6)
\right\} 
$
&
$\left[ + D_\beta  \frac{ 1}{ d (d+2)} \right]$
\end{tabular}
\end{minipage}
\end{table*}

\section*{Appendix C: One-Loop contributions to $D_\alpha, D_\beta$}

We here show how the diagram of Fig.\ \ref{fig2}c 
contribute to the disorder term, Fig.\ \ref{fig1}c.  
Due to the symmetry of the $D_{ijk}^{lmn}$ term in
Eq.\ (\ref{4}), the four possible types of diagrams in 
Fig.\ \ref{fig2}c contribute the same value.
 The result is
\begin{eqnarray}
L &=& 2^3 \left (- \frac{1}{2!} \right) \left( \frac{1}{2} \right)^2
\nonumber \\ && \times
\int d t_1 d t_2 
\int_{{\bf k}_1 {\bf k}_2 {\bf k}_3 {\bf k}_4} 
(2 \pi)^d \delta({\bf k}_1+ {\bf k}_2 + {\bf k}_3 + {\bf k}_4 )
\nonumber \\ && \times
\int d t_3 d t_4 
\int_{{\bf k}_5 {\bf k}_6 {\bf k}_7 {\bf k}_8} 
(2 \pi)^d \delta({\bf k}_5+ {\bf k}_6 + {\bf k}_7 + {\bf k}_8 )
\nonumber \\ && \times
\vert {\bf k}_1 + {\bf k}_2 \vert^{-y}
k_{2_j} k_{4_m} \hat{\bar b}_i^\perp({\bf k}_1,t_1)
      \hat b_k^\perp({\bf k}_2,t_1)
\nonumber \\ && \times
\vert {\bf k}_5 + {\bf k}_6 \vert^{-y}
k_{6_s} k_{8_v} \hat{\bar b}_r^\perp({\bf k}_5,t_3)
                   \hat b_w^\perp({\bf k}_8,t_4)
\nonumber \\ && \times
e^{-\nu k_6^2 (t_3-t_2)} \Theta(t_3-t_2) \hat \Pi_{lt}({\bf k}_6) 
(2 \pi)^d \delta({\bf k}_6 + {\bf k}_3)
\nonumber \\ && \times
e^{-\nu k_4^2 (t_2-t_4)} \Theta(t_2-t_4) \hat \Pi_{nu}({\bf k}_4) 
(2 \pi)^d \delta({\bf k}_4 + {\bf k}_7)
\nonumber \\ && \times
D_{ijk}^{lmn}
D_{rst}^{uvw}
\nonumber \\
&=&
\frac{-K_d}{2 \nu^2}
\bigg[
\frac{2}{d} \delta_{ms}  \delta_{lt} \delta_{nu}
\nonumber \\ &&
- \frac{2}{d (d+2)} \delta_{nu}\left(
  \delta_{ms} \delta_{lt}  + 
  \delta_{ml} \delta_{st}  + 
  \delta_{mt} \delta_{ls} 
\right)
\nonumber \\ &&
- \frac{2}{d (d+2)} \delta_{lt}\left(
  \delta_{ms} \delta_{nu}  + 
  \delta_{mn} \delta_{su}  + 
  \delta_{mu} \delta_{sn} 
\right)
\nonumber \\ &&
+ \frac{2}{d (d+2) (d+4)} M_{lmn}^{stu}
\bigg]
D_{ijk}^{lmn}
D_{rst}^{uvw}
\nonumber \\ && \times
     \int_{\Lambda/a}^\Lambda dq  q^{d-y-3}
\nonumber \\ && \times
\int d t_1 d t_2 
\int_{{\bf k}_1 {\bf k}_2 {\bf k}_3 {\bf k}_4} 
(2 \pi)^d \delta({\bf k}_1+ {\bf k}_2 + {\bf k}_3 + {\bf k}_4 )
\nonumber \\ && \times
\vert {\bf k}_1 + {\bf k}_2 \vert^{-y}
{{\bf k}_2}_j
{{\bf k}_4}_v 
\nonumber \\ && \times
      \hat {\bar b}_i^\perp ({\bf k}_1, t_1) \hat b_k^\perp ({\bf k}_2, t_1)
    \hat {\bar b}_r^\perp ({\bf k}_3, t_2) \hat b_w^\perp ({\bf k}_4, t_2)
\label{eqL1}
\end{eqnarray}
It is clear that to evaluate this expression, we need to evaluate
a term such as 
\begin{eqnarray}
\sum_{lmn, stu}  &&
k_{2_j} k_{4_v} \hat{\bar b}_i^\perp({\bf k}_1,t_1)
                             \hat b_k^\perp({\bf k}_2,t_1)
            \hat{\bar b}_r^\perp({\bf k}_3,t_2) \hat b_w^\perp({\bf k}_4,t_2)
\nonumber \\ && \times
D_{ijk}^{lmn}
D_{rst}^{uvw}
\left[ \delta_{lt} \delta_{ms} \delta_{nu} \right]
\label{eqL2}
\end{eqnarray}
Fourteen other terms need to be evaluated in order to calculate the
total contribution from Eq.\ (\ref{eqL1}).  These terms each contribute
in a form that can be cast as a contribution to
$D_\alpha$ and $D_\beta$ in Eq.\ (\ref{5}).  Shown in Table \ref{intL} are the terms and their contributions.
\begin{table*}[p]
\begin{minipage}[h]{6.00in}
\caption{Terms in the contributions from Fig.\ \ref{fig2}c, via
Eq.\ \ref{eqL2}.}
\label{intL}
\begin{tabular}{lll}
\hline
Term & Contribution to $D_\alpha$ & Contribution to $D_\beta$ \\
\hline
$\delta_{lt} \delta_{ms} \delta_{nu}$
&
$2 D_\alpha^2 +  4 D_\alpha D_\beta$
&
$4 D_\alpha^2 + (4+d) D_\alpha D_\beta + 3 D_\beta^2 $
\\
$ \delta_{lt} \delta_{mn} \delta_{su} $
&
0
&
$(4+2d) D_\alpha^2 + (6+2 d) D_\alpha D_\beta + 2 D_\beta^2 $
\\
$ \delta_{lt} \delta_{mu} \delta_{ns} $
&
$2 D_\alpha^2 +  4 D_\alpha D_\beta$
&
$4 D_\alpha^2 + (4+d) D_\alpha D_\beta + 3 D_\beta^2 $
\\
$ \delta_{lm} \delta_{ts} \delta_{nu} $
&
0
&
$4 D_\alpha^2 + (6+2 d) D_\alpha D_\beta + (2 + 2 d)D_\beta^2 $
\\
$ \delta_{lm} \delta_{nt} \delta_{su} $
&
0
&
$(4+2d) D_\alpha^2 + (4+3 d + d^2) D_\alpha D_\beta + (1+d) D_\beta^2 $
\\
$ \delta_{lm} \delta_{tu} \delta_{ns}  $
&
0
&
$(4+2d) D_\alpha^2 + (4+3 d + d^2) D_\alpha D_\beta + (1+d) D_\beta^2 $
\\
$ \delta_{ls} \delta_{mt} \delta_{nu} $
&
$2 D_\alpha^2 +  4 D_\alpha D_\beta$
&
$4 D_\alpha^2 + (4+d) D_\alpha D_\beta + 3 D_\beta^2 $
\\
$ \delta_{ls} \delta_{mn} \delta_{tu} $
&
0
&
$(4+2d) D_\alpha^2 + (6+2 d) D_\alpha D_\beta + 2 D_\beta^2 $
\\
$ \delta_{ls} \delta_{mu} \delta_{nt}  $
&
$2 D_\alpha^2 +  4 D_\alpha D_\beta$
&
$4 D_\alpha^2 + (4+d) D_\alpha D_\beta + 3 D_\beta^2 $
\\
$ \delta_{ln} \delta_{mt} \delta_{su} $
&
0
&
$(4+2d) D_\alpha^2 + (4+3 d + d^2) D_\alpha D_\beta + (1+d) D_\beta^2 $
\\
$ \delta_{ln} \delta_{ms} \delta_{tu} $
&
0
&
$(4+2d) D_\alpha^2 + (4+3 d + d^2) D_\alpha D_\beta + (1+d) D_\beta^2 $
\\
$ \delta_{ln} \delta_{mu} \delta_{st} $
&
0
&
$4 D_\alpha^2 + (6+2 d) D_\alpha D_\beta + (2 +2 d) D_\beta^2 $
\\
$ \delta_{lu} \delta_{mt} \delta_{ns} $
&
$2 D_\alpha^2 +  4 D_\alpha D_\beta$
&
$4 D_\alpha^2 + (4+2 d) D_\alpha D_\beta + 2 D_\beta^2 $
\\
$ \delta_{lu} \delta_{mn} \delta_{st}$
&
0
&
$4 D_\alpha^2 + 8 D_\alpha D_\beta + 4 D_\beta^2 $
\\
$ \delta_{lu} \delta_{ms} \delta_{nt}$
&
$2 D_\alpha^2 +  4 D_\alpha D_\beta$
&
$4 D_\alpha^2 + (4+2 d) D_\alpha D_\beta + 2 D_\beta^2 $
\end{tabular}
\end{minipage}
\end{table*}
Summing all these contributions to $D_\alpha$ and $D_\beta$,
we get the third and fourth flow equations of Eq.\ (\ref{8a}).

\bibliography{turbulence}

\end{document}